%% file: Real-time_DAQ_for_Image_Sensors.tex
\documentclass[journal]{IEEEtran}
\newcommand{\PreserveBackslash}[1]{\let\temp=\\#1\let\\=\temp}

\usepackage{color}
\usepackage{array}
\usepackage{multirow}
\usepackage{textcomp}  % correct micro (\textmu)
\usepackage{cite}
\usepackage{graphicx}
\usepackage{cite}
\usepackage{subfigure}
\usepackage[T1]{fontenc}

\graphicspath{{./fig/}}

\begin{document}

\title{Real-time Data Acquisition and Processing System for MHz Repetition Rate Image Sensors}

\author{
	A.~Mielczarek, D.~Makowski,~\IEEEmembership{Member,~IEEE}, A.~Napieralski,~\IEEEmembership{Senior Member,~IEEE},\\
	Ch.~Gerth, B.~Steffen

	\thanks{A.~Mielczarek, D.~Makowski, A.~Napieralski,
			are with the Lodz University of Technology, Poland (e-mail: amielczarek@dmcs.pl)}

	\thanks{Ch.~Gerth, B.~Steffen
			are with the Deutsches Elektronen-Synchrotron (DESY), Hamburg, Germany}
}

\maketitle

\input{src/Extended_Abstract}

\bibliographystyle{IEEEtran}
\bibliography{biblio}

\end{document}

%% file: src/Extended_Abstract.tex
\section{Introduction}

\begin{figure}[htb]
\centering
\includegraphics[width=0.5\textwidth]{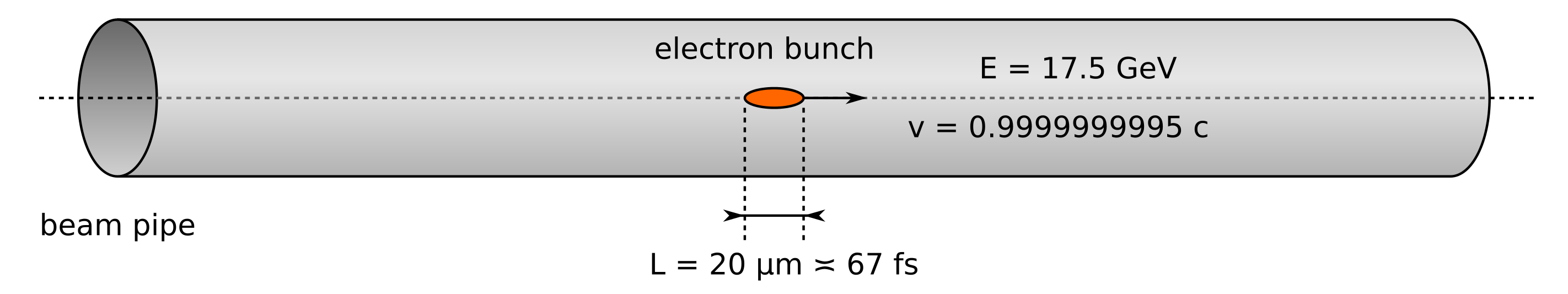}
\caption{Electron bunch propagating through XFEL beam pipe}
\label{fig:beam_pipe}
\end{figure}

One of the optimization goals of a particle accelerator is to reach the highest possible beam peak current.
For that to happen the electron bunch propagating through the accelerator should be kept relatively
short along the direction of its travel. As illustrated in Figure~\ref{fig:beam_pipe}, in case
of the European X-Ray Free-Electron Laser (E-XFEL) machine the length of a bunch is just around
20~\textmu{}m. The bunch leaves the accelerator with a nominal energy of 17.5~GeV. Therefore,
every single electron has the kinetic energy comparable to a falling snowflake (assuming typical
mass of 3~mg and speed of about 10~cm per second). The electrons move with the speed of 0.9999999995
of the speed of light in free space. The bunch passes through a particular point in space in just 67~fs. 

In order to obtain a better understanding of the beam composition it is crucial to evaluate
the electric charge distribution along these 20~\textmu{}m packets. Despite the scale of the challenge,
several methods of capturing a longitudinal charge profile of the bunch were developed.
Moreover, some of them allow for nondestructive real-time evaluation e.g. thanks to the
application of electro-optic crystals. The task of the Electro-Optic Detector (EOD) installed
at E-XFEL is to imprint the beam charge profile on the spectrum of light of a laser pulse.
The actual measurement of charge distribution is then extracted with a spectrometer
based on a diffraction grating~\cite{steffen2009compact}.

\section{Requirements}

The EOD spectrometer has to operate with repetition rate equal to the machine micro-bunching frequency.
In case of the E-XFEL accelerator this means reaching the frame rate of 4'500'000 frames per second.
To match the performance of the previously used Integrated Radiation Spectrometer, the detector shall
provide 256 readout channels with at least 14-bit resolution.

For the off-line machine performance analysis and tuning the device has to be able to capture and store
up to 2700 consecutive readouts and provide raw data to the host CPU with a repetition rate of at least
10~Hz. The samples are stored with 16-bit alignment, therefore, around 1.4~MB of memory is required for
storing data captured during a single bunch.

In order to maximize the noise immunity and reduce the number of required
external devices the bias voltage for the sensor has to be generated locally. In this field, the
silicon detector is more demanding of the two. It operates at the voltages between several tens
and around hundred Volts, which has to be remotely adjustable and free of noise.

\section{Hardware}

\begin{figure}[htb]
\centering
\includegraphics[width=0.4\textwidth]{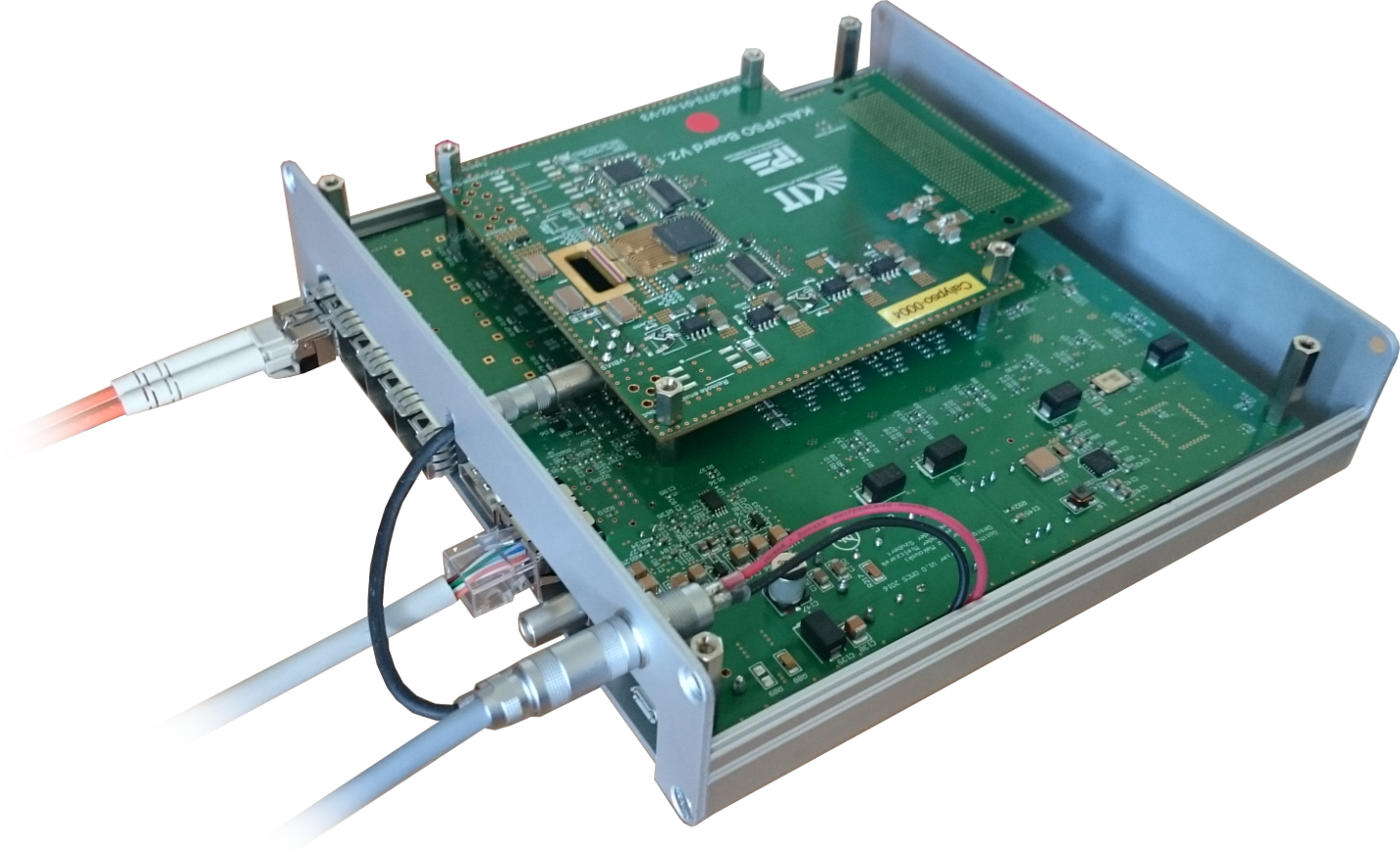}
\caption{Photograph and the module with KALYPSO detector}
\label{fig:hold_photo}
\end{figure}

The developed data acquisition and processing system is called the High-speed Optical Line Detector (HOLD).
It is a 1D image acquisition system which solves several challenges related to capturing, buffering, processing
and transmitting large data streams with use of the FPGA device. It implements a latency-optimized custom
architecture based on the AXI interfaces. The HOLD device is realized as an FPGA Mezzanine Card (FMC) carrier
with single High Pin-Count connector hosting the KIT KALYPSO detector~\cite{rota2017kalypso}. The assembled module
is presented in Figure~\ref{fig:hold_photo}.

\begin{figure}[htb]
\centering
\includegraphics[width=0.4\textwidth]{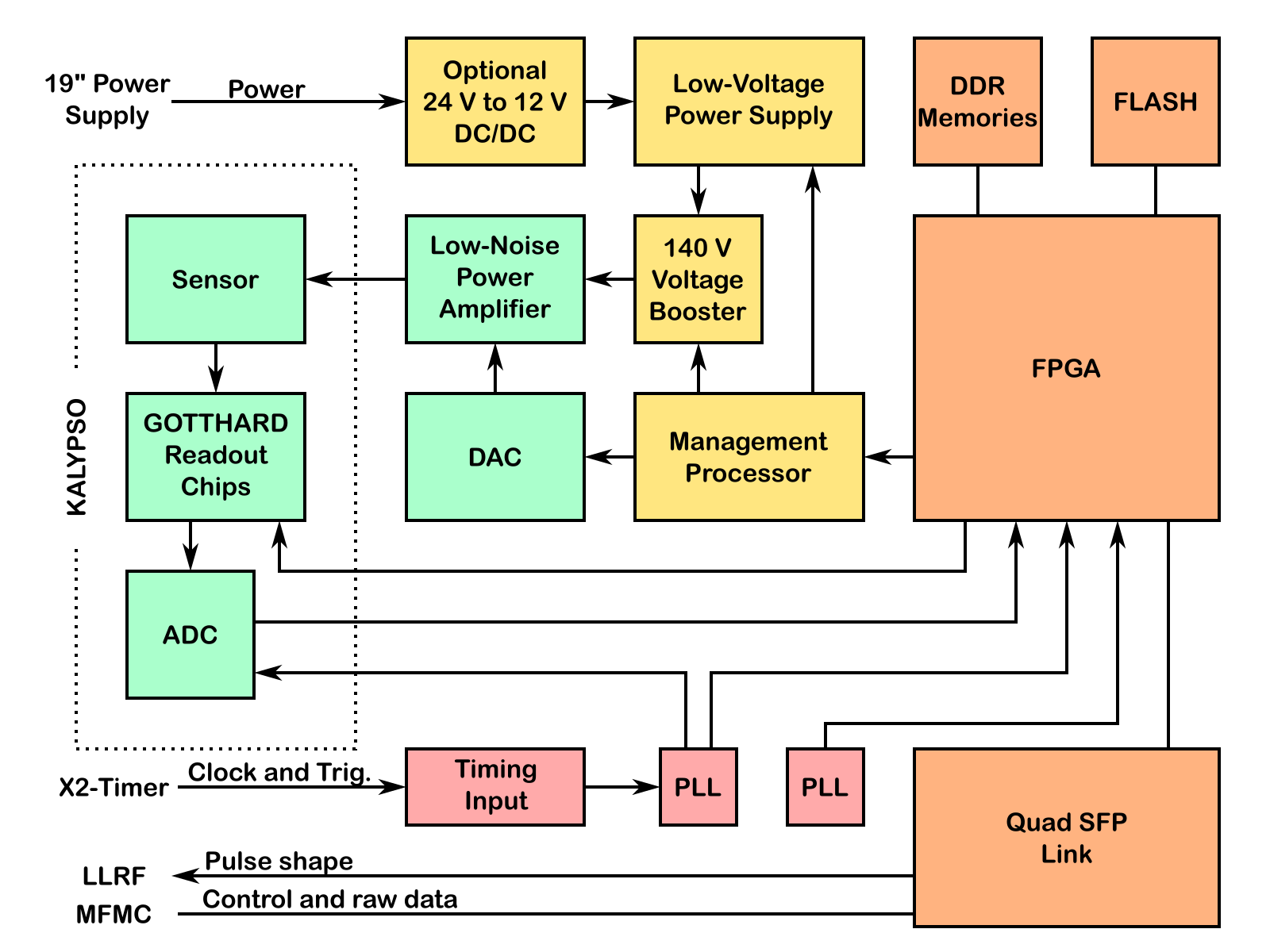}
\caption{Block Diagram of the Data Acquisition and Processing Module}
\label{fig:hold_hardware}
\end{figure}

The HOLD solution is built with a Xilinx 7-Series FPGA device which provides the processing power
and most of the high-performance interfaces. Its structure is presented in Figure~\ref{fig:hold_hardware}.
The FPGA is equipped with quad Multi-Gigabit optical link implemented with use of the Small Form-Factor
Pluggable (SFP) transceivers. This interface is used for control, raw data streaming as well as low-latency
communication channel to the Low-Level Radio Frequency (LLRF) system. This link is capable of providing
throughput up to 26~Gbps (four channels of up to 6.5~Gb/s). 

The timing is provided to HOLD from a dedicated timer module through a twisted-pair cable with modular
connector using the LVDS signaling. The clock signal is cleaned from jitter in one of the
Phase-Locked Loops (PLLs) and is then provided to FPGA and ADC on the KALYPSO mezzanine.
The FPGA utilizes this clock and one of the two triggers to control the GOTTHARD readout chips
of the detector board. The acquisition is constantly running, hence it can conveniently capture
dark frames and track the detector baseline. The second trigger is used to mark the start of
the next macro-pulse. For the purposes of development all the synchronization signals can be also
generated internally.

After the acquisition is enabled a configurable number of image frames is stored to the on-board
64-bit SDRAM memory. The number of frames usually does not exceed 2700, which is equivalent to
roughly 1.4~MB of data collected in less than 3~\textmu{}s. The HOLD device is equipped with 2~GB
of DDR3 memory, however only around 30~MB are used for the buffer.

The detector sensitivity depends on its bias voltage, therefore this parameter is remotely
controllable. This feature is implemented through two low-noise adjustable power supplies. One is
dedicated for InGaAs detector and provides voltages in range 0--10~V.
The other one is used with Si photo-diodes and can provide up to 120~V. In order to reduce the
power losses it is powered by an adjustable DC/DC boost converter, which tracks the output voltage
with a suitable margin. The power supply functionality is controlled by a dedicated micro-controller,
which communicates with FPGA over UART.

\section{Firmware}

\begin{figure}[htb]
\centering
\includegraphics[width=0.5\textwidth]{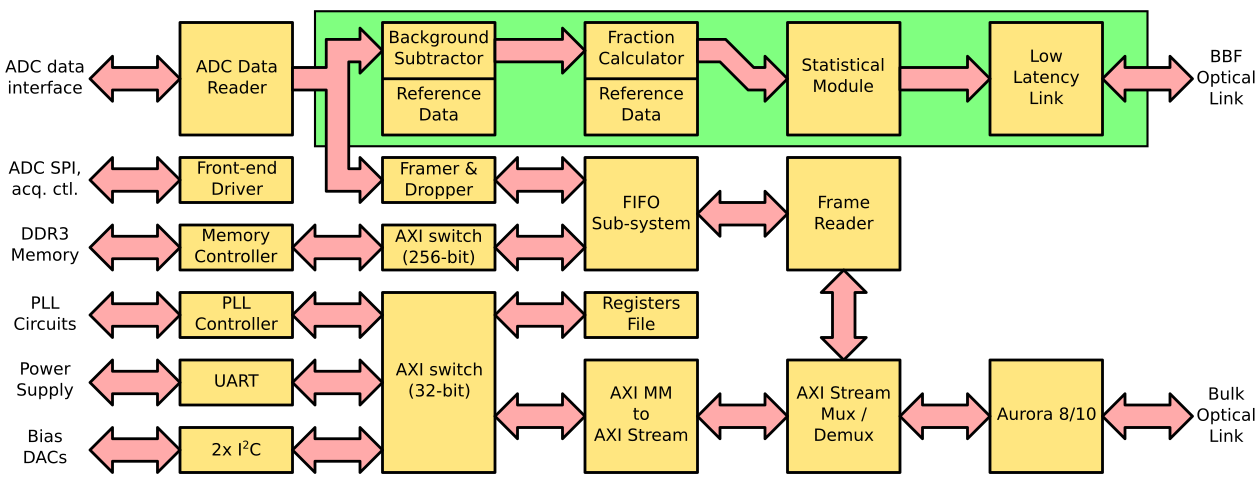}
\caption{Block Diagram of the HOLD FPGA firmware}
\label{fig:hold_firmware}
\end{figure}

The structure of the HOLD firmware is presented in Figure~\ref{fig:hold_firmware}. The design is
composed of a number of reusable blocks interfacing with use of Advanced Extensible Interface (AXI)
buses.

The first step of performing the data acquisition is to configure the ADC converter over an SPI interface. 
The ADC captures a new sample in every clock cycle, regardless of the readout chip state. To determine
which samples carry meaningful data, the reader module utilizes the "data valid" signal from the readout chip.

The GOTTHARD readout chip is, in principle, asynchronous -- the control signals directly influence its
operation, without synchronization to any clock. The front-end is driven by a Finite
State Machine (FSM), that is responsible for controlling the integrator, the sample-and-hold circuit
and the readout multiplexer. The readout sequence is triggered by an external signal from the timing module.
It is used to synchronize the GOTTHARD chips operation with arrival of the E-XFEL electron bunches.

The samples are stored in RAM and wait for clients to consume them. Just before transmission, each
packet is prepended by a framer with a simple header containing information on the number of acquired
image lines and the sequential number of the recorded bunch.

Simultaneously, the captured data can be also transferred to the other path, marked with green background,
which is focused on providing data for the BBF system. This part of the design is still under development.
Its purpose is to compute a number of parameters characterizing the bunch charge distribution, in particular:
position of the center-of-mass, lateral spread of the pulse and mean pixel readout. The calculated values
will be delivered to the LLRF system over a dedicated optical fiber.

\section{Conclusion}

The HOLD device is still under evaluation. The results will be presented in the full article.
The performed acquisitions prove that the electron bunches can indeed be measured by
the spectrometer set-up with use of the developed 1D camera. The solution presented in this paper
is probably one of the world fastest line cameras. Thanks to its custom architecture it is capable
of capturing at least 10 times more frames per second than fastest comparable commercially available devices.